\documentclass[journal]{IEEEtran}
\usepackage{xcolor}
\usepackage{amssymb}
\usepackage{environ}
\usepackage{graphicx}
\usepackage{subfigure}
\usepackage{algorithm}
\newcommand{\cN}{\mathcal{N}}

\renewcommand{\vec}[1]{\mathbf{#1}} 
\newtheorem{problem}{Problem}
\newtheorem{lemma}{Lemma}

\newtheorem{Proposition}{Proposition}

\ifCLASSINFOpdf

\else

\fi

\usepackage{amsmath}

\usepackage{algorithmic}

\begin{document}

\title{Distributed Coverage Control of Multi-Agent Networks with Guaranteed Collision Avoidance in Cluttered Environments} 

\author{Alaa~Z.Abdulghafoor,~ 
         Efstathios~Bakolas,~\IEEEmembership{Member,~IEEE}
\thanks{A. Abdulghafoor (graduate student) and E. Bakolas (Associate Professor) are with the Department of Aerospace Engineering and Engineering Mechanics, The University of Texas at Austin, Austin, Texas 78712-1221, USA. e-mails: alzabdulghafoor@utexas.edu; bakolas@austin.utexas.edu.\\ This research has been supported in part by NSF under award CMMI-1753687. \\ This is an extended version (including proofs) of the paper that has been accepted for inclusion in the 2021 Modeling, Estimation, and Control Conference (MECC 2021).}}

\maketitle

\begin{abstract}                
We propose a distributed control algorithm for a multi-agent network whose agents deploy over a cluttered region in accordance with a time-varying coverage density function while avoiding collisions with all obstacles they encounter. Our algorithm is built on a two-level characterization of the network. The first level treats the multi-agent network as a whole based on the distribution of the locations of its agents over the spatial domain. In the second level, the network is described in terms of the individual positions of its agents. The aim of the multi-agent network is to attain a spatial distribution that resembles that of a reference coverage density function (high-level problem) by means of local (microscopic) interactions of its agents (low-level problem). In addition, as the agents deploy, they must avoid collisions with all the obstacles in the region at all times. Our approach utilizes a modified version of Voronoi tessellations which are comprised of what we refer to as Obstacle-Aware Voronoi Cells (OAVC) in order to enable coverage control while ensuring obstacle avoidance. We consider two control problems. The first problem which we refer to as the high-level coverage control problem corresponds to an interpolation problem in the class of Gaussian mixtures (no collision avoidance requirement), which we solve analytically. The second problem which we refer to as the low-level coverage control problem corresponds to a distributed control problem (collision avoidance requirement is now enforced at all times) which is solved by utilizing Lloyd's algorithm together with the modified Voronoi tessellation (OAVC) and a time-varying coverage density function which corresponds to the solution of the high-level coverage control problem. Finally, simulation results for coverage in a cluttered environment are provided to demonstrate the efficacy of the proposed approach.
\end{abstract}


\IEEEpeerreviewmaketitle

\section{Introduction}
Increased attention has been paid recently to multi-agent networks which rely on distributed control algorithms to distribute roles and workload among themselves. In this work, a class of distributed coverage control problems is considered in which the agents of the multi-agent network are required to deploy over a given region while avoiding obstacles at all times such that their spatial distribution matches very closely a desired time-varying reference distribution.  

\textit{Literature Review}: The problem addressed herein lies under the category of deployment problems. Some examples include target tracking \cite{29} and area coverage \cite{6,10,13,27}. The objective of the deployment problem is to distribute the agents over a region in which a density function which is either static \cite{6} or time-varying \cite{10,13} describes the relative importance of each subset of the region of interest. A key problem in the deployment of multi-agent networks is how the agents maneuver while avoiding collisions among themselves and with obstacles in a cluttered domain of interest. Methods used for collision avoidance include leader-follower control strategies \cite{37}, potential field based methods \cite{47} and velocity obstacle methods \cite{36}. While extensive research has been made on the obstacle avoidance of agents in several applications, most algorithms depend on the robots' dynamics \cite{34}. Therefore, if the robot is subject to uncertainty or its dynamics are highly nonlinear, these algorithms may result in collisions with obstacles due to controller's poor performance. In addition, some of the proposed solution algorithms are not suitable for multi-agent systems as they do not guarantee safety among the agents. 

Another method for collision avoidance guarantees is the Voronoi-based coverage control; yet, this method guarantees collision among agents but not obstacle avoidance \cite{6}. If the latter method is modified, then effective obstacle collision can be guaranteed \cite{29}. Hence, in this work we utilize the Obstacle-Aware Voronoi Cell (OAVC) method \cite{29} that incorporates collision avoidance into path planning by generating a safe area around each agent. This method is based on the Voronoi-coverage control; however, to guarantee collision avoidance with obstacles, each generated Voronoi cell needs to be modified. One approach is to create buffered cells such that any agent located at the boundary of its cell does not collide with neighbours \cite{35,41}. However, \cite{35} use static weights to create the buffered cells which is useful if the offset between agents is known and the region of interest has no large obstacles. In contrast, in this paper we use the approach in \cite{29} which uses an algorithm that puts dynamic weights between the obstacles and agents such that the boundaries of the Voronoi cells are always tangent to the obstacles and never intersect or collide. 

\textit{Our Contributions}: In this paper we propose a two-level approach to the distributed control deployment problem in an environment with obstacles. The proposed approach aims to transport the multi-agent network from an initial Gaussian mixture (GM) distribution to a desired terminal GM distribution while guaranteeing collision safety and obstacle avoidance as the agents maneuver to achieve the goal spatial disposition. In particular, we show that there exists a path of GM connecting the initial and goal GMs which can be characterized in closed form in terms of the evolution of its components (mean, covariance and  mixing proportions) as functions of time. The GM solution acts as a reference density to the low-level problem, which corresponds to an obstacle avoidance coverage problem with time-varying density in which collision avoidance is supposed to be guaranteed at all times. To solve the latter problem, we propose a new distributed control algorithm which is a variation of Lloyd's algorithm in which OAVCs are utilized to guarantee obstacle avoidance as the agents move toward the locations that conform to the desired terminal distribution. In addition, our work considers a third problem that combines the low-level and high-level problems to indirectly steer the density of the team towards the desired reference density as the distributed control problem does not guarantee that. 
 
\textit{Outline:} In Section II, we formulate the three control problems, high-level, low-level and combined coverage control problems. In Section III, we present the solutions to the first two control problems. In Section IV, we present the algorithm that solves the combined coverage control problem based on the solutions to the high-level and the low-level control problems. In Section V, we present simulation results. Finally, in Section VI we conclude the paper with a summary of remarks. 

\section{Problem Formulation}
\subsection{Preliminaries}

\noindent 
Given two integers $j_a$ and $j_b$ with $j_a\leq j_b$, the discrete interval $j_a$ to $j_b$ is denoted by $[j_a, j_b]_d$. 
The space of $m$-dimensional vectors in $\mathbb{R}^{m}$ with non-negative components is denoted by $\mathbb{R}^{m}_{+}$. Moreover, the $m-1$ standard simplex is denoted by $\Delta^{m-1}$ where, $\Delta^{m-1}$ consists of all vectors $\lambda = [\lambda_1, \dots, \lambda_m ]^\mathrm{T} \in \mathbb{R}^m$ with $\lambda_j\geq 0, \forall j \in [1,m]_d$ and $\sum_{j=1}^m \lambda_j =1$.
To denote that the (symmetric) matrix $\boldsymbol{\Sigma}$ is positive definite, we write $\boldsymbol{\Sigma} = \boldsymbol{\Sigma}^{\mathrm{T}} > \mathbf{0}$.
We denote the probability density function (pdf) of a multi-variate ($k$-dimensional) Gaussian distribution $\mathcal{N}(\mu, \boldsymbol{\Sigma})$ with mean $\mu\in \mathbb{R}^{k}$ and covariance $\boldsymbol{\Sigma}\in \mathbb{R}^{k \times k}$ with $\boldsymbol{\Sigma} = \boldsymbol{\Sigma}^{\mathrm{T}} > \mathbf{0}$, as $\rho_{\mathcal{N}}(x; \mu, \boldsymbol{\Sigma})$  where
 \begin{equation}\label{g2}
\rho_{\mathcal{N}}(x; \mu, \boldsymbol{\Sigma}) := \frac{\exp\left(-\frac{1}{2}(x-
\mu)^{\mathrm{T}}\boldsymbol{\Sigma}^{-1}(x - \mu)\right)}{\sqrt{\mathrm{det}(\boldsymbol{\Sigma}) (2\pi)^k}}, 
 \end{equation}
where the determinant of $\boldsymbol{\Sigma}$ is denoted as $\mathrm{det}(\boldsymbol{\Sigma})$. 

Given a collection of $m$ Gaussian density functions $\mathcal{G} := \{ \rho_{\mathcal{N}}(x; \mu_j, \boldsymbol{\Sigma}_j): ~j\in[1,m]_d \}$), where $\mu_j \in \mathbb{R}^k$ and $\boldsymbol{\Sigma}_j = \boldsymbol{\Sigma}_j ^{\mathrm{T}} > \mathbf{0}$, for $j\in[1,m]_d$, and a vector of mixing proportions $\lambda \in \Delta^{m-1}$, we define the corresponding Gaussian mixture symbolized by $g(x;\mathcal{G}, \lambda)$ as follows:
 \begin{equation}\label{g3b}
 g(x;\mathcal{G}, \lambda) := \sum_{j=1}^{m}\lambda_{j} \rho_{\mathcal{N}}(x; \mu_j, \boldsymbol{\Sigma}_j).
 \end{equation}

Given a compact set $\Omega\subseteq \mathbb{R}^k$, and a point-set $P:=\{p_i \in \Omega: i\in [1, n]_d\}$, we refer the collection of sets $\mathcal{V}:=\{ \mathcal{V}_1, \dots, \mathcal{V}_n \}$
where
\begin{equation} \label{eq2}
\mathcal{V}_i := \{q \in \Omega: \| q-p_{i}\| \leq \| q-p_{j}\|,~ \forall i \neq j\},
\end{equation}
$\forall i \in[1,n]_d$, 
as the Voronoi Tessellation (VT) or Voronoi Partition (VP) \cite{21} of the set $\Omega$ generated by the point-set $P$, and the set $\mathcal{V}_i$ as the $i$-th Voronoi cell of the VP. The modified Voronoi cells (OAVC) for the $i$-th agents will be denoted as $\mathcal{A}_i$.

We denote by $\mathcal{O}=[o_1,\dots,o_j,\dots,o_{\bar{n}}]^{\mathrm{T}}$ the set of static obstacles represented as circles centred at $o_j$ with radius $r_j ~ \forall j \in[1,\bar{n}]_d$. We write $\tilde{\mathcal{O}}_{j}$ to denote the subset each obstacle occupies in the region $\Omega$ defined as : 
\begin{equation} \label{eq:Aobst}
\tilde{\mathcal{O}}_{j} := \{q \in \Omega: \| q-o_{j}\| \leq r_j,~ \forall j \in [1,\bar{n}]_d\},
\end{equation}
 while $\tilde{\mathcal{O}}$ denotes the total area occupied by all the obstacles, that is, $\tilde{\mathcal{O}}:=\bigcup_{j=1}^{\bar{n}}\tilde{\mathcal{O}}_j$.

Finally, given a continuous function $f(x): \Omega\subseteq \mathbb{R}^k \rightarrow \mathbb{R}$, where $\Omega$ is assumed to be a compact set, its spatial $\mathcal{L}_2$ norm is denoted by $\| f(x) \|_{\mathcal{L}_2}$ and defined as:
 \begin{equation}
     \| f(x) \|_{\mathcal{L}_2} := \left(\int_{\Omega} | f(x) |^2 \mathrm{d} x \right)^{1/2}.
 \end{equation}

\subsection{Problem Statement}
In this section, we will formulate the two obstacle avoidance coverage problems for a multi-agent network. 
\begin{problem}[High-level coverage control problem]\label{problem1}
\noindent \textbf{Given:} A 2-dimensional compact domain $\Omega$ over which $n$ agents are dispersed. Let the point-set $P_0:=\{p_i^0 \in \Omega: i\in [1, n]_d\}$ represent the agents' locations at time $t=t_0$, which is assumed to be a known initial Gaussian mixture distribution defined as $\phi_{P_0}(q):= g(q;\mathcal{G}_0, \lambda_0)$ where $\mathcal{G}_0 = \{ \rho_{\mathcal{N}}(q; \mu_{i,0}, \boldsymbol{\Sigma}_{i,0}): ~i\in[1,m_0]_d \}$, where $\mu_{i,0} \in \mathbb{R}^2$ and $\boldsymbol{\Sigma}_{i,0} = \boldsymbol{\Sigma}_{i,0}^\mathrm{T} > \mathbf{0}$ for all $i \in [1,m_0]_d$ and $\lambda_{0} \in \Delta^{m_0-1}$. Moreover,
we are given a terminal Gaussian mixture density distribution with $m$ components and corresponding mixing proportions, defined as $\phi_{P_f}(q):= g(q;\mathcal{G}_f, \lambda_f)$, where $\mathcal{G}_f = \{ \rho_{\mathcal{N}}(q; \mu_{j,f}, \boldsymbol{\Sigma}_{j,f}): ~j\in[1,m]_d \}$, where $\mu_{j,f} \in \mathbb{R}^2$ and $\boldsymbol{\Sigma}_{j,f} = \boldsymbol{\Sigma}_{j,f}^\mathrm{T} > \mathbf{0}$ for all $j \in [1,m]_d$ and $\lambda_{f} \in \Delta^{m-1}$. \\ 
\noindent \textbf{Goal:} Find a reference coverage density function $\phi(q,t):\Omega \times [0,\infty) \rightarrow  [0,\infty)$ such that: (i) $\phi(q,\cdot)$ is $\mathcal{C}^2$ over $\Omega$, (ii) $\phi(\cdot, t)$ is continuous for all $t\in [0,\infty)$, (iii) $\phi(q,t)\geq 0$, $\forall (q,t) \in \Omega \times [0,\infty)$ and (iv) the following boundary conditions are satisfied at $t=t_0$ and as $t\rightarrow \infty$:
\begin{align} 
\lim_{t\rightarrow t_0}\| \phi(q,t) - \phi_{P_0}(q)\|_{\mathcal{L}_2} & = 0, \label{phicon1} \\
\lim_{t\rightarrow \infty} \|\phi(q,t) - \phi_{P_f}(q)\|_{\mathcal{L}_2} & = 0. \label{phicon2}
\end{align} 
Thus, Problem~\ref{problem1} seeks for a density-path that will connect the initial density $\phi_{P_0}(q)$ with the terminal density $\phi_{P_f}(q)$.
\end{problem} 
Next, the low-level control problem is introduced. The objective of the latter problem is to find the individual inputs of the agents that will make the team of agents evolve such that its density will eventually converge (in the $\mathcal{L}_2$ sense) to the goal terminal density $\phi_{P_f}(q)$ while avoiding obstacles in a cluttered region. For this purpose, a dynamic coverage optimization problem which utilizes the solution to (Problem~\ref{problem1}) as its time-varying reference density is proposed. We consider a group of $n$ homogeneous mobile robots distributed over $\Omega$, having a set of $\tilde{n}$ static obstacles $\mathcal{O}$ centered at $o_j$ for all  $j\in[1,\tilde{n}]_d$. The motion of the mobile robots is described as follows: 
\begin{equation}\label{eq:motion}
    \dot{p_{i}}(t) = u_{i}(t),~~~p_i(t_0)=p_i^0,~~~i\in[1,n]_d,
\end{equation}
where $p_{i}(t)$ = $[x_{i}(t) \; \; y_{i}(t)]^{\mathrm{T}}$ is the position of $i$th agent in $\Omega$ at time $t$ and $u_{i}(t) = [u_{x_{i}}(t) \; \;u_{y_{i}}(t)]^{\mathrm{T}}$ is its velocity vector (control input). 

 \begin{problem}[Low-level coverage control problem] \label{problem2}
 \noindent \textbf{Given:} The set of static obstacles $\mathcal{O}$, the set of initial locations of the $n$ agents $P_0:=\{p_i^0 \in \Omega: i\in [1, n]_d\}$ such that $p_i(t_0) \notin \mathcal{\tilde{O}}$ for all $i\in[1,n]_d$ and a reference density function $\phi(q,t)$ that solves Problem~\ref{problem1}. In addition, a configuration cost function that measures how well the agent $p_{i}$ $\in$ $\Omega$ is positioned which is defined as:
\begin{equation} \label{eq:H}
H(p,t;\mathcal{A})= \sum_{i=1}^{n} \int_{\mathcal{A}_{i}}\|q-p_{i}\|^{2} \phi(q,t) d\Omega.
\end{equation}
 \noindent \textbf{Goal:}{ Design distributed control algorithms to control the motion of agents in order to minimize the locational cost defined in (\ref{eq:H}). In other words,
 \begin{align}\label{eq5}
 \indent \indent \;\;\; & \underset{p}{\text{minimize}}
 && H(p,t;\mathcal{A}) \\ \nonumber
 & \text{subject to}
 && \dot{p}(t)=u \;\; \mathrm{as} \;\;  t \rightarrow \infty,
 \end{align}}
where $p:=[p^{\mathrm{T}}_{1}, \dots, p^{\mathrm{T}}_{n}]^{\mathrm{T}}$.
 \end{problem}
Due to the fact that neither the high-level coverage control problem nor the low-level problem result independently in a satisfactory solution to the multi-agent control problem that we are trying to solve comprehensively, we will propose a combined version of these two problems, which will be solved practically. 
To be specific, Problem~\ref{problem1} seeks a density path that will connect the initial GM density to a desired terminal GM density but overlooks the agents' dynamics, the obstacles and the individual control inputs that these agents need to apply for this density path to be realized. On the other hand, Problem~\ref{problem2} seeks for the individual control inputs that will account for collision avoidance and solve a locational optimization to help the network obtain the goal spatial distribution specified in the  high-level problem, yet with no assurance that the solution to the latter problem will allow the density of the agents to approach the goal density. Thus, the previous reasons prompt us to consider a third problem that seeks for the computation of the individual inputs of the agents (similar to Problem~\ref{problem2}) that will steer the density of the agents towards the desired terminal density (similar to Problem~\ref{problem1}).

To formulate the third problem, we consider an approximation of the team's probability density by a Gaussian mixture (density of the actual distribution of the network's agents), which we refer to as the \textit{fine} approximation and denote as $\phi_{\mathrm{team}}$. For the computation of the fine approximation at each instant of time $t$, the set of positions of the agents $P(t):=\{p_i(t) \in \Omega: i\in [1, n]_d\}$ is fitted into a Gaussian mixture density: 
\begin{align} \label{teamhat}
\phi_{\mathrm{team}}(q,t) &= g(q;\mathcal{G}_{\mathrm{team}}(t), \lambda_{\mathrm{team}}(t)) \nonumber\\
&=\sum_{j=1}^{m}\lambda_{j,\mathrm{team}}(t) \rho_{\mathcal{N}}(q; \mu_{j,\mathrm{team}}(t), \boldsymbol{\Sigma}_{j,\mathrm{team}}(t)),
\end{align} 
with the use of, for instance, the Gaussian Mixture Model Likelihood Optimization (GMMLO) algorithm. 
Having the point-set $P(t)$ and a positive integer $m$ as input to the GMMLO algorithm, the output of the latter algorithm will be a Gaussian mixture consisting of $m$ Gaussian mixtures, which form the collection $\mathcal{G}_{\mathrm{team}}(t):= \{ \rho_{\mathcal{N}}(q; \mu_{j,\mathrm{team}}(t), \boldsymbol{\Sigma}_{j,\mathrm{team}}(t)):~j\in[1,m]_d \}$, with corresponding mixing proportion vector $\lambda_{\mathrm{team}}(t)$.
\begin{problem}[Combined coverage control problem]\label{problem3} 
\noindent \textbf{Given:}
Let $\phi_{\mathrm{team}}(q,t_0):=g(q;\mathcal{G}_{\mathrm{team}}(t_0), \lambda_{\mathrm{team}}(t_0))$ be the fine approximation of the Gaussian mixture density distribution of the population based on the dispersion of the agents at time $t\geq t_0$. \\
\noindent \textbf{Goal:} Find the individual control inputs $u_i$, for $[1,n]_d$, that will steer the agents emerging from the point-set $P_0$, whose motion is described by \eqref{eq:motion}, to the terminal destinations forming a point-set $P_f:=\{p_i^f \in \Omega: i\in [1, n]_d\}$, that corresponds to (approximately) a sample deduced from a probability distribution with density $\phi_{P_f}$. In other words, we seek to enforce the following limiting behaviors:
\begin{align} 
\|\phi_{\mathrm{team}}(q,t) - \phi_{P_0}(q)\|_{\mathcal{L}_2} &\rightarrow 0~~\text{as}~~ t\rightarrow t_0,\label{phiteam condition0} \\
\|\phi_{\mathrm{team}}(q,t) - \phi_{P_f}(q)\|_{\mathcal{L}_2} &\rightarrow 0~~\text{as}~~ t\rightarrow \infty.\label{phiteam conditionf}
\end{align}
\end{problem}

\section{Proposed Solution and Analysis}
\subsection{Solution of the High-Level Coverage Control Problem}
In this section, we will introduce a solution to Problem~\ref{problem1}. For this purpose, consider a function $\phi(q,t)$ which corresponds to a time-varying Gaussian mixture reference coverage density defined as:
 \begin{align}\label{eq11}
 \phi(q,t):= g(q;\mathcal{G}(t), \lambda(t))&=\sum_{j_1=1}^{m_0}\lambda_{j_1,0}(t)\phi_{j_1,0}(q,t_0) \nonumber \\
   &+\sum_{j_2=1}^{m}\lambda_{j_2}(t)\phi_{j_2}(q,t), 
 \end{align}
 for all $(q,t) \in \Omega \times [0,\infty)$, where $\phi_{j_1,0}(q,t)$ and $\phi_{j_2}(q,t)$ are given by 
\begin{align}\label{eq12}
 \phi_{j_1,0}(q,t_0)&= \rho_{\mathcal{N}}(q; \mu_{{j_1},0}, \boldsymbol{\Sigma}_{{j_1},0})), \quad j_1\in [1,m_0]_d,\\
\phi_{j_2}(q,t)&= \rho_{\mathcal{N}}(q; \mu_{j_2}(t), \boldsymbol{\Sigma}_{j_2}(t)), \quad j_2\in [1,m]_d
\end{align}
where $\mu_{{j_1},0} \in \Omega$ and $\boldsymbol{\Sigma}_{{j_1},0}=\boldsymbol{\Sigma}_{{j_1},0}^\mathrm{T} >0$ for $j_1\in [1, m_0]_d$ denote the means and covariances at $t=t_0$ and $\mu_{j_2}(t) \in \Omega$ and $\boldsymbol{\Sigma}_{j_2}(t) = \boldsymbol{\Sigma}_{j_2}(t)^{\mathrm{T}} > \mathbf{0}$, for $j_2\in [1, m]_d$, symbolize the (time-varying) means and covariances, respectively, of the Gaussian densities that determine, together with the vector of mixing proportions $\lambda_{j_2}(t)$, the Gaussian mixture density path at each time $t\geq t_0$.

Our goal is to find a time-varying Gaussian mixture $\phi(q,t):= g(q;\mathcal{G}(t), \lambda(t))$, which satisfies the following boundary (limiting) conditions:
\begin{align}\label{eqphi0phifinal}
& \lim_{t \rightarrow t_0} \| g(q;\mathcal{G}(t), \lambda(t)) - g(q;\mathcal{G}_0, \lambda_0)\|_{\mathcal{L}_2} = 0, \\ 
& \lim_{t \rightarrow \infty}\|g(q;\mathcal{G}(t), \lambda(t)) - g(q;\mathcal{G}_f, \lambda_f) \|_{\mathcal{L}_2} = 0, \label{eqphi0phifinal2}
\end{align}
where $g(q;\mathcal{G}_0, \lambda_0)$, $g(q;\mathcal{G}_f, \lambda_f)$,  $\mathcal{G}_0$ and $\mathcal{G}_f$ are defined as in Problem~1. Next, we provide a closed-form solution to Problem~1.
\begin{Proposition}\label{prop1}
Let us consider the collection of Gaussians $\mathcal{G}(t) = \{\rho_{\cN}(q; \mu_{j_1,0}, \boldsymbol{\Sigma}_{j_1,0})\} \cup \{\rho_{\cN}(q; \mu_{j_2}(t), \boldsymbol{\Sigma}_{j_2}(t)): j_1\in[1,m_0]_d ~and~ j_2\in[1,m]_d \}$,
where
\begin{align}\label{eq:muj}
 \mu_{j_2}(t) & = \mu_{j_2,f} + (\mu_{j_1,0}-\mu_{j_2,f})\exp(a(t-t_{0})), \\
\boldsymbol{\Sigma}_{j_2}(t) & =\boldsymbol{\Sigma}_{j_1,0}^{-1/2} \big[\exp(b(t-t_{0}))\boldsymbol{\Sigma}_{j_1,0} \nonumber \\
&~~~\quad~~~~ +(1-\exp(b(t-t_{0})) \nonumber \\
&~~\qquad~~ \times (\boldsymbol{\Sigma}_{j_1,0}^{1/2}\boldsymbol{\Sigma}_{j_2,f}\boldsymbol{\Sigma}_{j_1,0}^{1/2})^{1/2}\big]^{2}\boldsymbol{\Sigma}_{j_1,0}^{-1/2},\label{eq:Sigmaj} 
\end{align}
for $j_1 \in [1, m_0]_d$, $j_2 \in [1, m]_d$ and $t\geq 0$, where $a<0$, $b<0$. In addition, let $\lambda(t):= [\lambda_{j_1,0}(t), \dots, \lambda_1(t), \dots, \lambda_m(t)]^{\mathrm{T}}$ for $j_1\in[1,m_0]_d$ with
\begin{align}
\lambda_{j_1,0}(t)&=\lambda_{j_1,0}~e^{\alpha (t-t_0)}\label{eq:lambda0} ~~j_1\in[1,m_0]_d, \\
\lambda_{j_2}(t)&=\lambda_{j_2,f}\big(1-\sum_{j_1=1}^{m_0}\lambda_{j_1,0}(t)\big),~~j_2\in[1,m]_d,\label{eq:lambda}
\end{align}
for all $t\geq t_0$, where $\alpha<0$  and $\lambda_{f} \in \Delta^{m-1}$ is the vector of mixing proportions corresponding to $\mathcal{G}_f = \{ \rho_{\cN}(q; \mu_{j_2,f}, \boldsymbol{\Sigma}_{j_2,f}): ~j_2\in[1,m]_d \}$. Also, $\sum_{j_1=1}\lambda_{j_1,0} =1$.
Then, the function $\phi: \Omega \times [0,\infty) \rightarrow [0, \infty)$, where $\phi(q,t) := g(q;\mathcal{G}(t), \lambda(t))$ satisfies the boundary conditions (\ref{phicon1}) and (\ref{phicon2}) and thus solves Problem ~\ref{problem1}.
\end{Proposition} 

\begin{IEEEproof}
From (\ref{eq:muj}) and \eqref{eq:Sigmaj}, it follows readily that 
\begin{align} 
\lim_{t\rightarrow t_0} \mu_{j_2}(t) & =\mu_{{j_1},0}, & \lim_{t\rightarrow t_0} \boldsymbol{\Sigma}_{j_2}(t) & = \boldsymbol{\Sigma}_{j_1,0}, \label{C12} \\
\lim_{t\rightarrow \infty} \mu_{j_2}(t) & =\mu_{j_2,f}, & \lim_{t\rightarrow \infty} \boldsymbol{\Sigma}_{j_2}(t) & = \boldsymbol{\Sigma}_{j_2,f} \label{C34},
\end{align}
for $j \in [1,m]_d$. In addition, (\ref{eq:lambda0}) and (\ref{eq:lambda}) imply
\begin{align}
\lim_{t\rightarrow t_0}\lambda_{j_1,0}(t)&= \lambda_{j_1,0}, 
&\lim_{t\rightarrow \infty} \lambda_{j_1,0}(t)&= 0, \label{Clambda0f} \\
 \lim_{t\rightarrow t_0}\lambda_{j_2}(t)&= 0, 
 &\lim_{t\rightarrow \infty} \lambda_{j_2}(t)&=\lambda_{j_2,f},  \label{C2lambda0f}
\end{align} 
Equations (\ref{eq:lambda0}) and (\ref{eq:lambda}) also imply that $\lambda(t) \in \mathbb{R}^{m+m_0}_{+}$. In addition, we will show that $\| \lambda(t)\|_1=1$, for all $t \geq 0$, where
\begin{align*}
    \| \lambda(t) \|_1 &= \sum_{j_1=1}^{m_0}|\lambda_{j_1,0}(t)|+ \sum_{j_2=1}^m|\lambda_{j_2}(t)| 
\end{align*}
and we will conclude that $\lambda(t) \in \Delta^{m_0+m-1}$, for all $t\geq 0$. Indeed, we have  
\begin{align}
    \| \lambda(t) \|_1 & = \sum_{j_1=1}^{m_0}\lambda_{j_1,0}(t)+ \sum_{j_2=1}^m \lambda_{j_2}(t) \nonumber \\
    &=\sum_{j_1=1}^{m_0}\lambda_{j_1,0}\exp(\alpha(t-t_0)) \nonumber \\
    &~~~+ (\sum_{j_2=1}^m \lambda_{j_2,f})(1-\sum_{j_1=1}^{m_0}\lambda_{j_1,0}\exp(\alpha(t-t_0)) )  \nonumber \\
    & = 1,
\end{align}
for all $t \geq t_0$, where in the last equality, we have used the fact that $\sum_{j=1}^{m}\lambda_{j_2,f} =1$. Therefore, $\lambda(t) \in \Delta^{m_0+m-1}$, for all $t \geq t_0$. We conclude that the collection of Gaussians $\mathcal{G}(t) = \{\rho_{\cN}(q; \mu_{j_1,0}, \boldsymbol{\Sigma}_{j_1,0}): j_1\in[1,m_0]_d\} \cup \{\rho_{\cN}(q; \mu_{j_2}(t), \boldsymbol{\Sigma}_{j_2}(t)): j_2\in[1,m]_d \}$ with corresponding mixing proportions $\{\lambda_{j_1,0},\dots,\lambda_{m_0,0},\lambda_1,\dots, \lambda_m \}$ determine a Gaussian mixture for all $t \geq t_0$, whose density $ \phi(q,t)$ satisfies equations (\ref{eq11}).
Next, we show that the density $\phi(q,t)$, which is defined in \eqref{eq11}, satisfies the boundary conditions (\ref{phicon1}) and (\ref{phicon2}). From \eqref{C12}-\eqref{C34} and \eqref{Clambda0f}-\eqref{C2lambda0f}, we conclude
\begin{align} 
\lim_{t\rightarrow t_0}\| \phi(q,t) - \phi_{P_0}(q)\|_{\mathcal{L}_2} & = 0, \label{C3} \\
\lim_{t\rightarrow \infty} \|\phi(q,t) - \phi_{P_f}(q)\|_{\mathcal{L}_2} & = 0, \label{C4}
\end{align} 
where in the last derivation we have used the fact that $\phi_{P_0}(q) = \sum_{j_1=1}^{m_0} \lambda_{j_1,0} \phi_{j_1,0}(q,t)$ and $\phi_{P_f}(q) = \sum_{j_2=1}^m \lambda_{j_2,f} \phi_{j_2}(q,t)$.This completes the proof.
\end{IEEEproof}

\subsection{Solution of the Low-Level Coverage Control Problem}
To solve Problem (\ref{problem2}) we design a distributed control algorithm that will make the agents track the time-varying centroids of their Voronoi cells and asymptotically converge to them while avoiding collisions with obstacles and among themselves. The solution approach proposed to address the low-level coverage control problem depends on a variation of Lloyd's algorithm for the case of a time-varying density. The approach will include 1) computation of the modified Voronoi tessellations of the spatial domain comprised of the Obstacle Aware Voronoi Cells (\cite{29}) generated by the current locations of the agents to avoid any obstacles in the domain, 2) characterization of the agent's individual control inputs based on the information obtained from their own Voronoi cells. The Lloyd's approach includes generating the modified Voronoi tessellations and computing the mass $M_{i}$ and the centroids $C_{i}$ of the $i$th OAVC and iteratively modifying the agents' positions $p_{i}$ to the centroids $C_{i}$. 

 \subsubsection{Obstacle Aware Voronoi Cell}
 To create a modified Voronoi cell for obstacle collision avoidance, we adopt the approach proposed in \cite{29}. When the agents calculate the Voronoi boundaries between other agents and themselves, they use the standard Voronoi tessellation method \cite{19} which allows collision avoidance among the agents. However, to account for the presence of bigger obstacles, the agents will dynamically assign weights for each obstacle such that the boundaries of the modified Voronoi cells are tangent to the obstacle boundaries. The modified Voronoi cell is referred to as the “Obstacle-Aware Voronoi Cell” (OAVC), defined by: 
\begin{align} \label{OAVC}
\mathcal{A}_i &:= \{q \in \Omega: \| q-p_{i}\|^2 \leq \| q-o_j\|^2-w_{ij},~ o_j\in \mathcal{O} \nonumber \\ &\mathrm{and} ~ 
\| q-p_{i}\|^2 \leq \| q-p_k\|^2, ~ \forall ~k~ \in[1,n]_d \neq i\},
\end{align}
where $w_{i,j}$ is the dynamic weight that forces the boundary lines of the Voronoi cell to be tangent to the obstacles defined as 
\begin{equation}
w_{ij} := 2r_j\|p_{i}-o_j\|-\|p_i-o_j\|^2,
\end{equation} \label{weight}
for all $j \in[1,\tilde{n}]_d$ and for all $i \in[1,n]_d$. 
Therefore, with the dynamic weight in the OAVC, the largest possible convex cell for the agent around static obstacles will be created. In addition, by maintaining a convex cell, the agents will be able to utilize  move-to-centroid distributed controllers and satisfy asymptotic convergence to the centroids  while they guarantee collision avoidance with other agents or obstacles. 

Given a density function $\phi(q,t): \Omega \times [0,\infty) \rightarrow \mathbb{R} $, the mass $M_i(\mathcal{A}_i,t)$ and centroid $C_i(\mathcal{A}_i,t)$ of the $i$th OAVC are defined as follows:
 \begin{subequations}
 \begin{align} 
& M_i(\mathcal{A}_i,t) := \int_{\mathcal{A}_{i}}\phi(q,t) d\Omega, \label{eq3} \\
& C_i(\mathcal{A}_i,t) := \frac{1}{M_i(\mathcal{A}_i,t)} \int_{\mathcal{A}_i} \phi(q,t) q d\Omega, \label{Ci}
 \end{align} 
  \end{subequations}
 where their derivatives with respect to time are computed as follows:
 \begin{subequations}
 \begin{align} \label{eq4}
 \dot{M_i}(\mathcal{A}_i,t)& = \int_{\mathcal{A}_i}\dot{\phi}(q,t) d\Omega, \\
 \dot{C_i}(\mathcal{A}_i,t) & =\frac{1}{M_i(\mathcal{A}_i,t)}\Big( \int_{\mathcal{A}_{i}}q\dot{\phi}(q,t) d\Omega \\ & ~~\quad~~ -\dot{M}_i(\mathcal{A}_i,t) C_i(\mathcal{A}_i,t) \Big).
 \end{align} 
\end{subequations}
Thus, in order for the agents to achieve asymptotic tracking of the time-varying centroids of their Voronoi cells, a feedback controller that will ensure that $H(p,t;\mathcal{A})$ will decrease along the agents' trajectories must be designed. In particular we have, 
 \begin{align}\label{eq15}
 \frac{\partial H(p,t;\mathcal{A})}{\partial p_{i}}&= \int_{\mathcal{A}_{i}}\frac{\partial\|q-p_{i}\|^{2} }{\partial p_{i}}\phi(q,t) d\Omega \nonumber \\ 
 &=  \int_{\mathcal{A}_{i}}-2(q-p_{i})^{T}\phi(q,t) d\Omega 
 \end{align}
 and by expanding the above expression and using (\ref{eq3}) and (\ref{Ci}) we obtain 
 \begin{align}\label{eq16}
 \frac{\partial H(p,t;\mathcal{A})}{\partial p_{i}}= 2M_{i}(p_{i}-C_{i})^{\mathrm{T}}, \ \forall \ i \in [1, n]_d,
 \end{align}
 where $M_{i}>$ 0 and $\frac{\partial H(p,\mathcal{A})}{\partial p_{i}}= 0$ when $p_{i} = C_{i}$. 
 We can compute the derivative of $H(p,t;\mathcal{A})$ as
 \begin{align}\label{eq17}
 \frac{d{H(p,t;\mathcal{A})}}{dt} &= \frac{\partial H(p,t;\mathcal{A})}{\partial t} + \frac{\partial H(p,t;\mathcal{A})}{\partial p} \dot{p} \nonumber \\
 &=\sum_{i=1}^{n} \int_{\mathcal{A}_{i}}\|q-p_{i}\|^{2} \frac{\partial\phi(q,t)}{\partial t} d\Omega \nonumber \\ 
 & \quad+ 2M_{i}(p_{i}-C_{i})^{T} \; \dot{p}_{i}.
 \end{align}
 Thus, to ensure that $\dot{H}(p,\mathcal{A},t)$ is negative semi-definite, we propose the following feedback control law:
 \begin{align}\label{eq18}
u_i(t,p_i;\mathcal{A}_i) := -\left(k_{0}+\frac{k_1}{M_{i}}\int_{\mathcal{A}_{i}}\|q-p_{i}\|^{2}d\Omega\right)(p_{i}-C_{i}).
 \end{align}

\subsection{Analysis of the Low-Level Solution and Results}
Following~\cite{22}, it can be shown that when $\frac{\partial\phi(q,t)}{\partial t}$ is bounded from above, the existing upper bound can be used for the design of a controller that will solve Problem \ref{problem2}.
\begin{Proposition}\label{prop2}
Let us assume that there exists $A_{1}\geq 0$ such that
\begin{equation}\label{eq20}
A_{1}\geq \sup_{t\geq 0,q\in \Omega} \left| \frac{\partial\phi(q,t)}{\partial t}\right|.
\end{equation}
Moreover, let us assume that the proportional gain $k_{1}$ of the controller given in (\ref{eq18}) satisfies the following inequality
\begin{equation}\label{eq22}
k_{1}\geq\frac{A_{1}}{2\|(p_{i}-C_{i})\|^{2}}.
\end{equation}
Then, the controller given in (\ref{eq18})
makes the time derivative of the locational cost $H(p,t;\mathcal{A})$ which is defined in (\ref{eq:H}), negative-semi definite along the trajectories of the agents of the network, that is, 
\begin{equation*}
    \dot{H}(p,t; \mathcal{V}) \leq 0,~~~~\forall ~~ t\geq 0.
\end{equation*}
\end{Proposition} 

\begin{IEEEproof}
In view of (\ref{eq18}), we have
\begin{align}\label{eq19}
\dot{H}(p,\mathcal{V},t)&=\sum_{i=1}^{n} \int_{\mathcal{V}_{i}}\|q-p_{i}\|^{2} \frac{\partial\phi(q,t)}{\partial t}d\Omega \nonumber \\
& -2k_{0}M_{i}\|(p_{i}-C_{i})\|^{2}\nonumber \\
&-2k_{1}\|(p_{i}-C_{i})\|^{2}\int_{\mathcal{V}_{i}}\|q-p_{i}\|^{2}d\Omega 
\end{align}
using (\ref{eq20}) it follows by inspection of (\ref{eq19}) that
\begin{align}\label{eq21}
\dot{H}(p,t;\mathcal{V})&\leq\sum_{i=1}^{n}[A_{1}-2k_{1}\|(p_{i}-C_{i})\|^{2}]\nonumber \\ &~~~~ \times\int_{\mathcal{V}_{i}}\|q-p_{i}\|^{2}d\Omega\nonumber \\ 
&~~~~ -2k_{0}M_{i}\|(p_{i}-C_{i})\|^{2}.
\end{align}
Therefore, if the proportional gain $k_1$ satisfies (\ref{eq22}), then $A_{1}-2k_{1}\|(p_{i}-C_{i})\|^{2}\leq 0$. Thus, \eqref{eq21} implies that $\dot{H}(p(t),\mathcal{V},t) \leq 0$, $\forall ~ t\geq 0$ and the proof is complete.
\end{IEEEproof}

\begin{Proposition}\label{prop3}
The system of mobile agents located at $p_{i}(t)$ at time $t \geq 0$, for $i\in [1, n]_d$, driven by the feedback control law (\ref{eq18}) will converge to the trajectories of the time-varying Voronoi centroids $C_{i}$ which evolve based on the reference time-varying density $\phi(q,t)$.
\end{Proposition} 
Next, we will prove that the mobile robots will track and eventually converge to the (moving) centroids of their Voronoi cells (the latter correspond to critical points of the locational cost $H(p,t;\mathcal{V})$ for the given spatial domain of interest $\Omega$). The proof of convergence we will provide is based on the following lemma. 

\begin{lemma}[Barbalat]\label{lemma1}
	Consider a function $V(p,t)$ which satisfies the following properties:\\
	(1) V(p,t) is lower bounded, \\
	(2) $\dot{V}(p,t)$ is negative semi-definite, \\
	(3) $\dot{V}(p,t)$ is uniformly continuous in time \\
	Then, $\lim_{t\to\infty} \dot{V}(p,t) = 0$
\end{lemma}

\begin{IEEEproof}
To prove the statement, we will show that the following candidate Lyapunov function $V(p,t)=H(p,t;\mathcal{A})$ satisfies all the three properties of Lemma 1. \\
(1) Since the time-varying Gaussian mixture density $\phi(q,t)$ is assumed to be strictly positive, then it is clear that the locational cost $H(p,t;\mathcal{A})$ (\ref{eq:H}) is positive definite. Hence $V(p,t)=H(p,t;\mathcal{A})>0$ and thus $V$ is lower bounded. \\
(2) In light of (\ref{eq20}) and (\ref{eq21}), $\dot{V}(p,t)=\dot{H}(p,\mathcal{A},t)$ (\ref{eq19}) would be negative semi-definite. ($\dot{V}(p,t)=\dot{H}(p,\mathcal{A},t)\leqslant0$).\\
(3) To prove that $\dot{V}(p,t)$ is uniformly continuous in time, it suffices to prove that the second time derivative of the Lyapunov function $\ddot{V}(p,t)=\ddot{H}(p, t; \mathcal{A})$ is bounded. 
We have, 
\begin{align}
\ddot{V}(p,t)&=\ddot{H}(p,\mathcal{A},t)\nonumber\\
&=\frac{d}{dt}\left[\frac{\partial H(p,t;\mathcal{A})}{\partial t} + \frac{\partial H(p,t;\mathcal{A})}{\partial p} \dot{p}\right] \nonumber \\
&=\sum_{i=1}^{n} \int_{\mathcal{A}_{i}}\|q-p_{i}\|^{2} \frac{d}{dt}\left[\frac{\partial\phi(q,t)}{\partial t}\right] d\Omega \nonumber  \\
&\quad +2\dot{M}_{i}(p_{i}-C_{i})^{T}\dot{p}_{i}+2M_{i}(\dot{p}_{i}-\dot{C}_{i})^{T}\dot{p}_{i}\nonumber\\
&\quad + 2M_{i}(p_{i}-C_{i})^{\mathrm{T}}\ddot{p}_{i}, \label{eq25}
\end{align}
it can be observed that all the terms on the right hand side of (\ref{eq25}) are finite and bounded, which suggests that $\ddot{V}(p,t)$ is finite and bounded. Also in view of the fact that all the hypotheses of Barbalat's Lemma are satisfied, it follows that  $\lim_{t\to\infty} \dot{V}(p,t) = 0$, thus concluding that the agents asymptotically converge to their respective time-varying Voronoi centroids. 
\end{IEEEproof}

\begin{Proposition}\label{prop4}
For the system of mobile agents positioned at $p_i(t_0) \notin \mathcal{\tilde{O}}$, with each having an OAVC $\mathcal{A}_i$ (\ref{OAVC}), the feedback control law (\ref{eq18}) guarantees that $p_i(t) \not= p_k(t)~ \forall~ k\in [1,n]_d\not=i$ and $p_i(t) \notin \mathcal{\tilde{O}}, \forall ~ t\geq t_0$.
\end{Proposition} 
\begin{IEEEproof}
By the definition of $\mathcal{A}_i$ (\ref{OAVC}), each agent's cell is convex. The centroid of any convex set lies inside the convex hull of the vertices of $\mathcal{A}_i$. Therefore, since the agents move to the centroids of their cells, they will only move within the collision free safe-area avoiding any collisions. In addition, because $\mathcal{A}_i(t_0)$ does not intersect with any obstacle or other agents, then all subsequent calculation of $\mathcal{A}_i$ will not intersect $\mathcal{O}$. 
\end{IEEEproof}

\section{Solution to the Combined Coverage Control Problem}
This section presents the solution to Problem \ref{problem3} by iteratively combining the solutions to Problems 1 and 2. The proposed approach is based on estimating the fine approximation of the team's density at different instants of time which form a finite increasing sequence $\{\tau_k\}_{k=0}^K$, where $\tau_0=t_0$ and $K$ is a positive integer; in particular, $\tau_k= t_0 + k \Delta \tau$ for $k \in \{0,\dots, K\}$, where $\Delta\tau>0$ is a given time-step. At each time instant $\tau_k$, we compute the fine approximation $\phi_{\mathrm{team}}(q,\tau_k)$ which is determined by the mean $\mu_\mathrm{team}(\tau_k)$ and covariance $\boldsymbol{\Sigma}_\mathrm{team}(\tau_k)$ of the locations of the team of agents at that time, which are computed by the GMMLO at $t=\tau_k$.
Therefore, $\phi_{\mathrm{team}}(q,\tau_k) = g(q;\mathcal{G}_{\mathrm{team}}(t), \lambda_{\mathrm{team}}(t))$. Then, we set $t_0 := \tau_k$, $\mu_{j_1,0} := \mu_{j_2,\mathrm{team}}(\tau_k)$, and $\boldsymbol{\Sigma}_{j_1,0} := \boldsymbol{\Sigma}_{j_2,\mathrm{team}}(\tau_k) ~\forall j_1 \in [1,m_0]_d \mathrm{and} ~\forall j_2 \in [1,m]_d$; subsequently, we solve Problem~1 to acquire a path of Gaussian mixtures, denoted by $\phi_k(q,t)$, where the means and covariances of its Gaussian components and the corresponding mixing proportions are given by \eqref{eq:muj}-\eqref{eq:lambda}. In other words, the solution to the high-level problem is updated at time $t = \tau_k$ resulting in a reference density $\phi_k(q,t)$, for $t\geq \tau_k$.
Subsequently, Problem \ref{problem2} (low-level problem) will be solved using $\phi_k(q,t)$ for all $t \in [\tau_k, \tau_{k+1})$ for $k\in [0, K-1]_d$ to acquire a distributed controller for the multi-agent network according to \eqref{eq18}. By applying the updated distributed control law for $t \in [\tau_k, \tau_{k+1})$, the agents will begin maneuvering in the plane in such a way that the fine approximation of the team's density $\phi_{\mathrm{team}}(q,t)$ will try to track the reference density $\phi_k(q,t)$. At time $t = \tau_{k+1}$, we set $t_0: =\tau_{k+1}$ and we repeat the steps described above to get a new reference density and execute the new distributed controller for time $t \in [t_{k+1}, t_{k+2})$. At $t= \tau_K$ we stop updating the reference density and the function $\phi_K(q,t)$ will correspond to the reference density that we will be utilized for all times $t \geq \tau_K$. Therefore, the proof of convergence given in Proposition \ref{prop3} holds true, for $\phi(q,t) := \phi_K(q,t)$ and $t_0:=\tau_K$; thus, the agents will converge to the time-varying Voronoi centroids whose evolution is dictated by the reference density $\phi_K(q,t)$. Also, it is expected that the fine approximation of the team's density will conform (or ideally, get close) to the desired one as $t \rightarrow \infty$.

Next, we describe the exact steps of the proposed algorithm based on the pseudocode provided in Algorithm~1. At the first step (line 8), we compute the fine approximation of the team's density $\phi_{0}(q) =\phi_{\mathrm{team}}(q,t_0)$. At the second step (line 9), we compute the OAVCs $\mathcal{A}(P)$ generated by the point-set $P(t)$, which is comprised of the locations of the agents at time $t$. At the third step (lines 12, 15), we initialize the mean vectors $\mu_{{j_1},0}=\mu_{{j_2},\mathrm{team}}(\tau_k)$, covariance matrices $\boldsymbol{\Sigma}_{{j_1},0}=\boldsymbol{\Sigma}_{{j_2},\mathrm{team}}(\tau_k)$ and the mixing proportions $\lambda_{{j_1},0}=\lambda_{{j_2},\mathrm{team}}(\tau_k)$ of the team based on $\phi_{\mathrm{team}}(q,\tau_k)$. At the fourth step (lines 13, 16), we compute the mean vectors $\mu_{{j_2}}(t)$ defined in \eqref{eq:muj}, the covariance matrices $\boldsymbol{\Sigma}_{{j_2}}(t)$ defined in \eqref{eq:Sigmaj}, and the mixing proportions $\lambda_{{j_1},0}(t)$ and $\lambda_{{j_2}}(t)$ defined in \eqref{eq:lambda0}-\eqref{eq:lambda}, which determine  the reference Gaussian mixture density $\phi_k(q,t)$, which is defined in \eqref{eq11}. At the fifth step (lines 21-24), we compute the reference density $\phi_k(q,t)$, mass $M_{i}$ and the centroids $C_{i}$ defined in (\ref{eq3}) and (\ref{Ci}), respectively, for each OAVC $\mathcal{A}_i$. At the sixth step (line 25), we compute the controller $u_i$ (\ref{eq18}) that is meant to drive the agents to follow $\phi_k(q,t)$ (\ref{eq11})-(\ref{eq12}). At the seventh and last step (line 28), the agents' positions are updated accordingly after the execution of the control input computed at the previous step. In this way, we obtain new density functions $\hat{\phi}_{\mathrm{team}}(q,t)$ and $\phi_{\mathrm{team}}(q,t)$ which will be used in the next iteration $t=\tau_{k+1}$ to update the reference density $\phi_k(q,t)$. All the previously described steps will be iterated until the convergence error $e_{\max}$ becomes less than a threshold $\epsilon>0$ (chosen a priori).

\begin{algorithm} 
	\caption{Agents' deployment algorithm}
	\label{tb1}
	\begin{algorithmic}[1]
		\STATE\textbf{Inputs:}{ domain dimensions $\Omega$, number of agents $n$, dymanic's time step $dt$, time step $\Delta\tau$, $\mathcal{G}_{f}$, the centers of the obstacles $O$ and the radius of each obstacle $r_j$ $a$, $b$, $\alpha$, $c$, controller gains ($k_{0}$,$k_{1}$)}, $\epsilon$
	    \STATE Let $k=0$
	     \STATE Let $t=t_0$
	      \STATE Let $e_{\max}= 1$
	    \WHILE{$e_{\max} > \epsilon$}
		\STATE Let $\tau_k=t$
		\STATE Let $\tau_{k+1}=t_0+(k+1)\Delta\tau$
		\STATE Find $\phi_{\mathrm{team}}(q,\tau_k)$
		\STATE Generate the $\mathcal{V}(P)$ 
		\STATE Let $\mu_{{j_2}}(\tau_0)=\mu_{{j_1},\mathrm{team}}(\tau_0)$ and $\boldsymbol{\Sigma}_{{j_2}}(\tau_0)= \boldsymbol{\Sigma}_{{j_1},\mathrm{team}}(\tau_0)$ 
	   	\IF{$\tau_k \leq \tau_K$}
	    	\STATE Let $\mu_{{j_1},0}=\mu_{{j_2},\mathrm{team}}(\tau_k)$, $\boldsymbol{\Sigma}_{{j_1},0}= \boldsymbol{\Sigma}_{{j_2},\mathrm{team}}(\tau_k)$ and $\lambda_{{j_1},0}=\lambda_{{j_2},\mathrm{team}}(\tau_k)$
	    	\STATE Compute $\lambda_{{j_1},0}(\tau_{k+1},\tau_k)$, $\lambda_{{j_2}}(\tau_{k+1},\tau_k)$, $\mu_{{j_2}}(\tau_{k+1},\tau_k)$ and $\boldsymbol{\Sigma}_{{j_2}}(\tau_{k+1},\tau_k)$  
		\ELSE
			\STATE Let $\mu_{{j_1},0}=\mu_{{j_2},\mathrm{team}}(\tau_K)$, $\boldsymbol{\Sigma}_{{j_1},0}= \boldsymbol{\Sigma}_{{j_2},\mathrm{team}}(\tau_K)$ and $\lambda_{{j_1},0}=\lambda_{{j_2},\mathrm{team}}(\tau_K)$
			\STATE Compute $\lambda_{{j_1},0}(\tau_{k},\tau_K)$, $\lambda_{{j_2}}(\tau_{k},\tau_K)$, $\mu_{{j_2}}(\tau_{k},\tau_K)$ and $\boldsymbol{\Sigma}_{{j_2}}(\tau_{k},\tau_K)$  
	    \ENDIF
        \STATE $k=k+1$ (Increment $k$)
         \STATE $e_i=0$
		\FOR{$i=1:n$} 
			\STATE Compute $\phi_k(q,\tau_k)$
		\FOR{$t_2=\tau_k:c dt:\tau_{k+1}$}
		    \STATE Compute $M_{i}(\mathcal{A}_{i},\tau_k)$
		    \STATE Then compute $C_{i}(\mathcal{A}_{i},\tau_k)$
		    \STATE Compute $\dot{p}_{i}=u_{i}$
		\ENDFOR
		\STATE  Compute $e_{i}=\|p_i-C_i\|$
		\STATE Update $p_{i}$
		\ENDFOR
		\STATE $t=t+\Delta\tau$ (increment $t$)
	    \STATE Let $e_{\max}=\max_{i}e_i$
		\ENDWHILE
	\end{algorithmic}
\end{algorithm}

\section{Numerical Simulations}
In this section we present numerical simulations. The simulations were conducted using $n = 10$ agents which we have initially positioned according to a Gaussian mixture distribution with $m_0=2$ but restricted in $\Omega : = [-20,20]\times[-20,20]$. The first and the second Gaussian mixands of the Gaussian mixture have mean vectors equal to $[3 \;12]^{\mathrm{T}}$ and $[12 \;14]^{\mathrm{T}}$, covariance matrices equal to $[0.2~-0.6;-0.6~3]$ and $[10.5~-0.5;-0.5~2]$ and mixing proportions equal to 0.3 and 0.7 respectively. 
\begin{figure}[tbh!]
	\centering
	\subfigure[$t=0$]{\includegraphics[width=4.3cm]{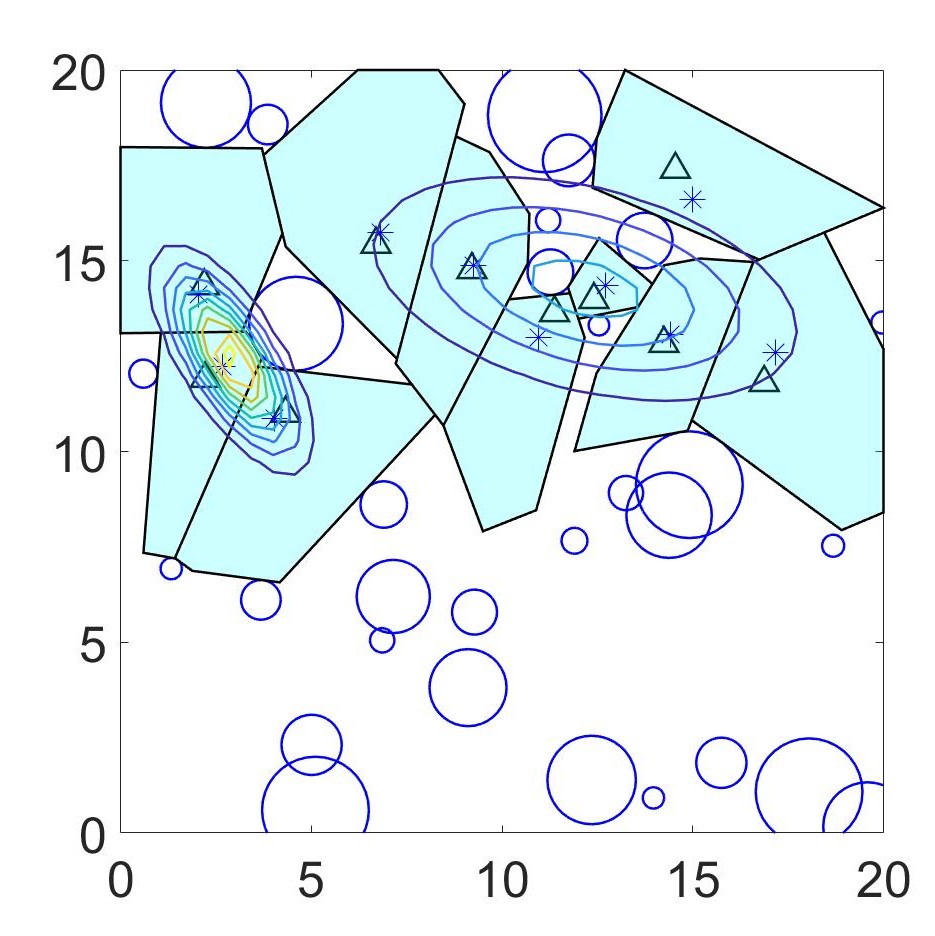}} 
	\subfigure[$t = 20\Delta\tau$]{\includegraphics[width=4.3cm]{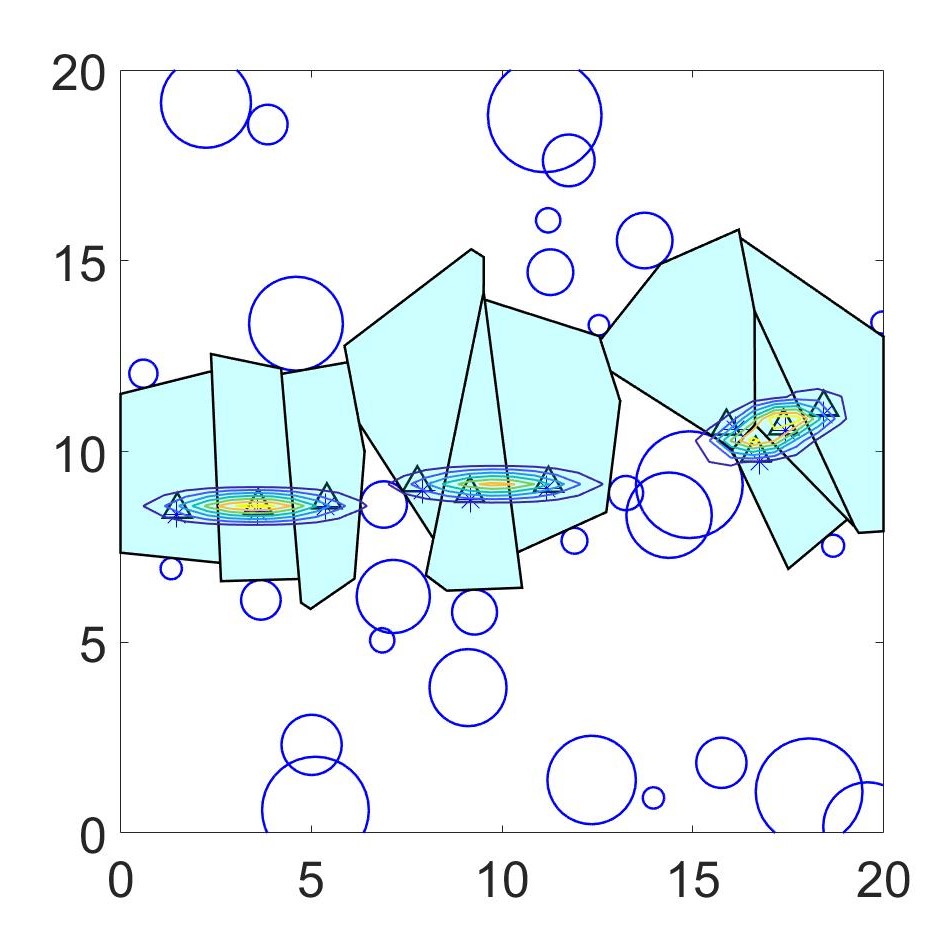}} 
	\subfigure[$t = 60\Delta\tau$]{\includegraphics[width=4.3cm]{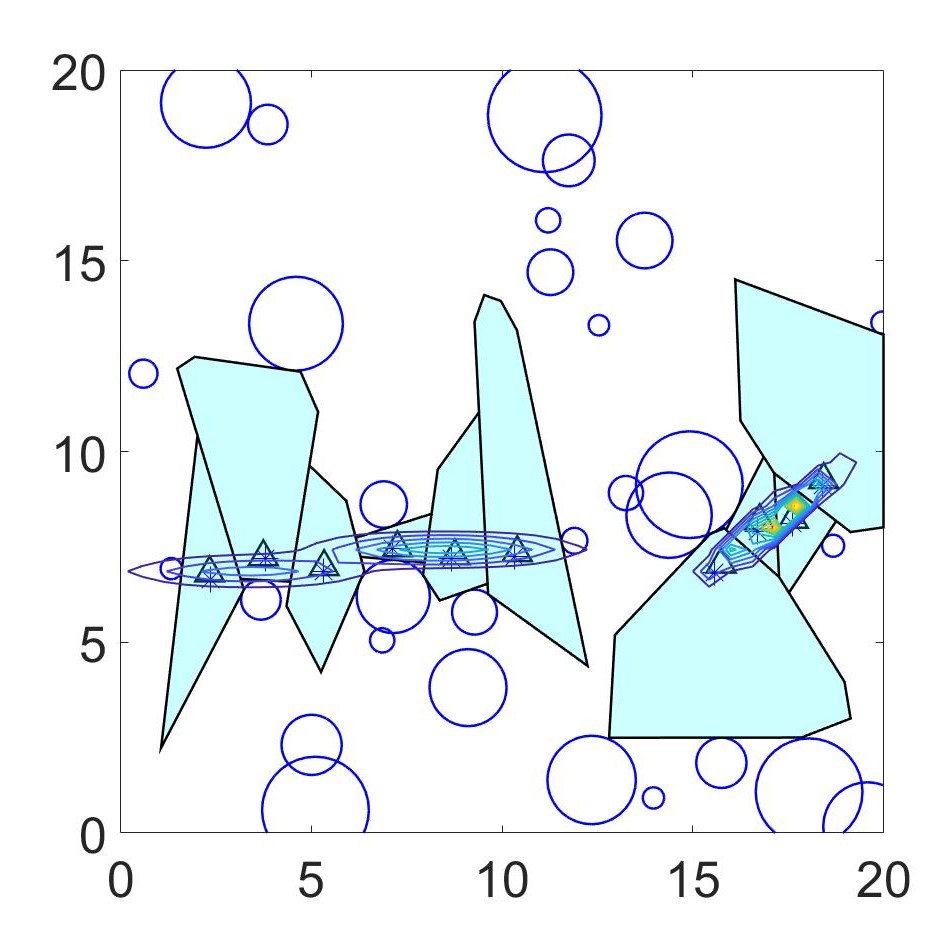}} 
	\subfigure[$t = 120\Delta\tau$]{\includegraphics[width=4.3cm]{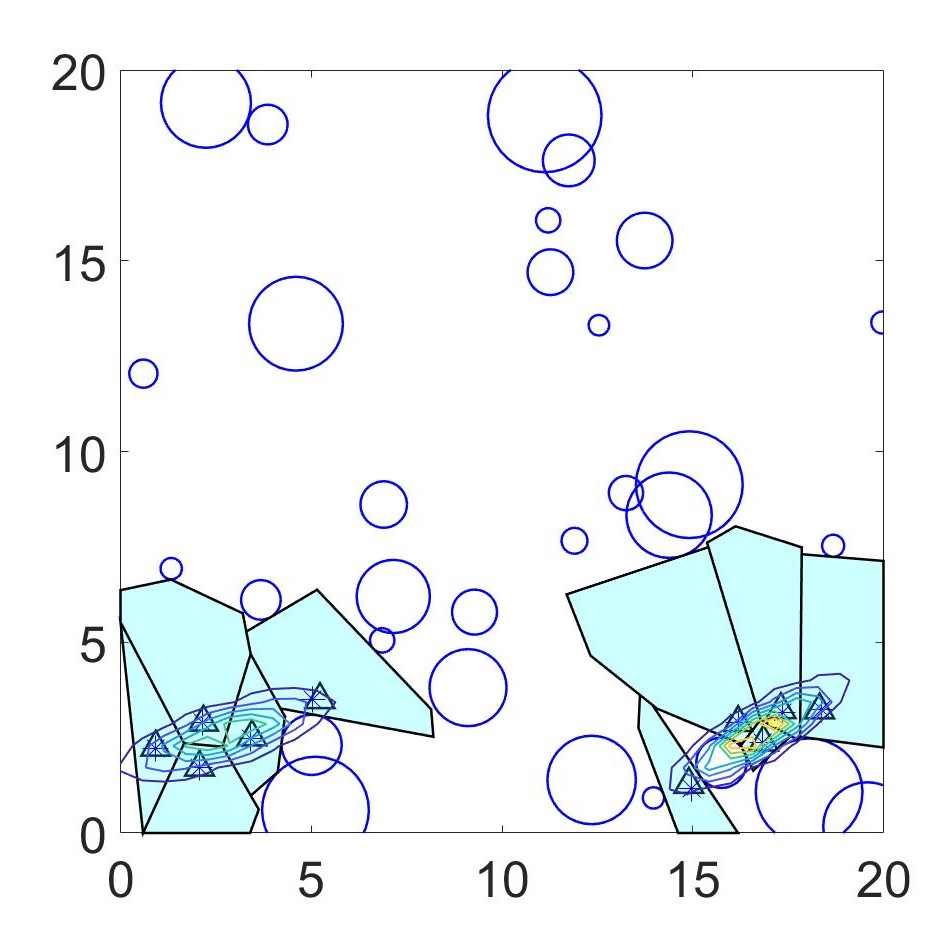}} 
	\caption{ Voronoi partitions at time (a) $t=0$, (b) $t=20\Delta\tau$, (c) $t=60\Delta\tau$, (d) $t=120\Delta\tau$. The agents are denoted as the small triangles, the centroids as the small stars, the obstacles as the circular disk and the OAVC as the blue polygons. The colored circular lines reflect the density contours of the fine approximation of the agents' density.}
	\label{Vcells}	
\end{figure}

Fig.~\ref{Vcells} shows the contours of the fine approximation of agents' density. Furthermore, Fig.~\ref{Vcells}, demonstrates the modified Voronoi cells (OAVC) displayed as filled blue polygons created by the agents as they move to track the time-varying Voronoi centroids corresponding to the reference time-varying coverage density $\phi(q,t)$ while avoiding collisions with the set of static obstacles $\mathcal{O}$ (displayed as circular disks) scattered in the domain $\Omega$. In addition, it can be shown that initially there is a difference in the positions of the agents (denoted as triangles) and the centroids of their OAVC (denoted as stars) as they are trying to track $\phi(q,t)$, yet this difference decreases as they asymptotically converge to their centroids as $t\rightarrow \infty$.

\begin{figure}[tbh!]
\centering
\begin{center}
\includegraphics[width=5.4cm]{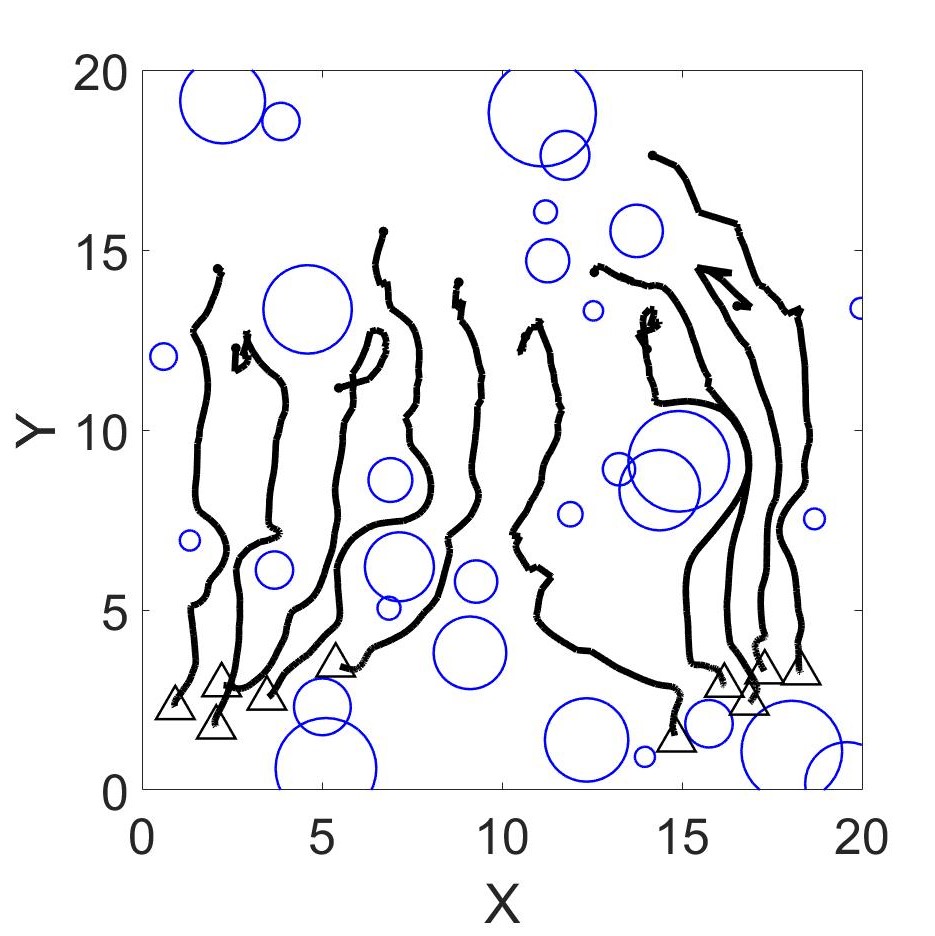}    
\caption{The trajectories of the agents as they move to track the reference density $\phi(q,t)$.} 
\label{path}
\end{center}
\end{figure}

Fig.~\ref{path} shows the trajectories of the agents that are driven by the control law (\ref{eq18}) which converge to their critical configurations. Moreover, we observe in Figure \ref{Vcells}(d) and Fig.~\ref{path} that the density of the team of agents as the latter approach their limit locations conforms with the desired goal Gaussian mixture density comprised of two Gaussians with mean vectors $(1,1)$ and $(17,1)$, covariance matrices $[0.7~0.2; 0.2~0.5]$ and $[0.8 ~ 0.2; 0.2 ~0.4]$ and mixing proportions equal to $0.5$ for both (limit of the reference density as $t\rightarrow \infty$). 

\begin{figure}[tbh!]
\begin{center}
\includegraphics[width=\linewidth]{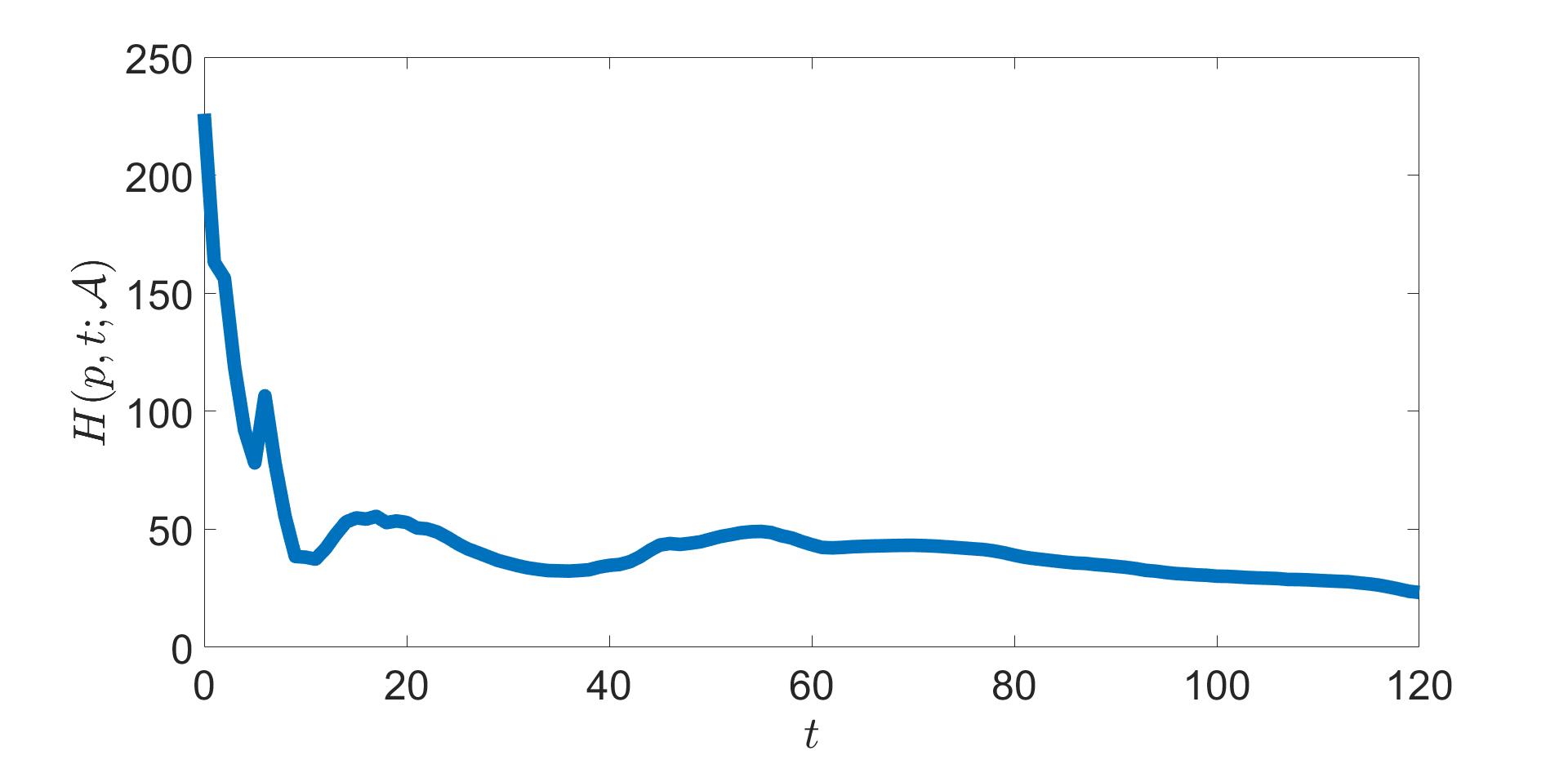}    
\caption{Locational cost $H(p,t;\mathcal{A})$ versus time.} 
\label{cost}
\end{center}
\end{figure}

Fig.~\ref{cost} illustrates the performance of the proposed controller in terms of minimizing the locational cost $H(p,t;\mathcal{V})$.

\begin{figure}[tbh!]
\centering
	\subfigure[]{\includegraphics[width=\linewidth]{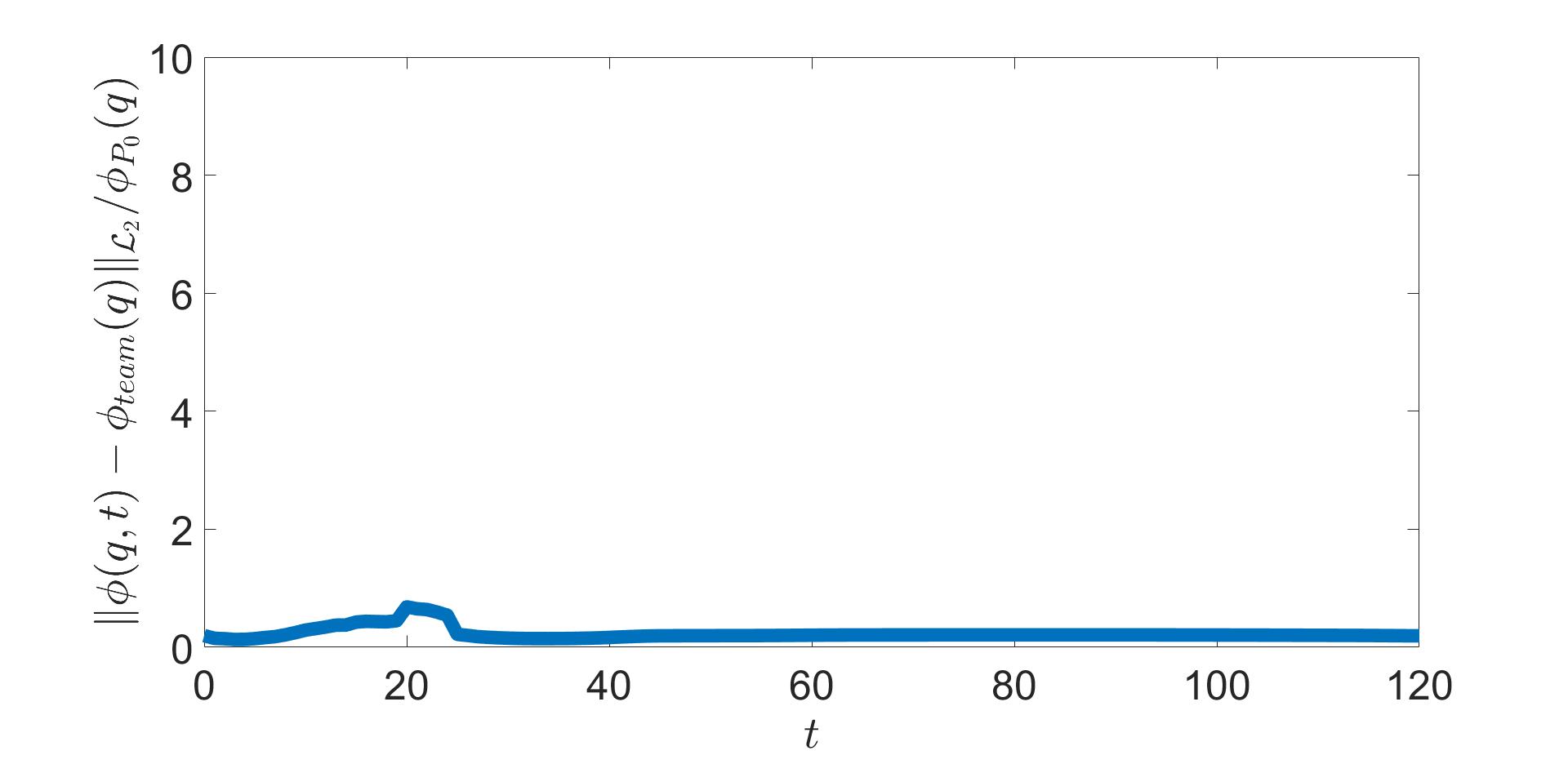}} 
	~
	\subfigure[]{\includegraphics[width=\linewidth]{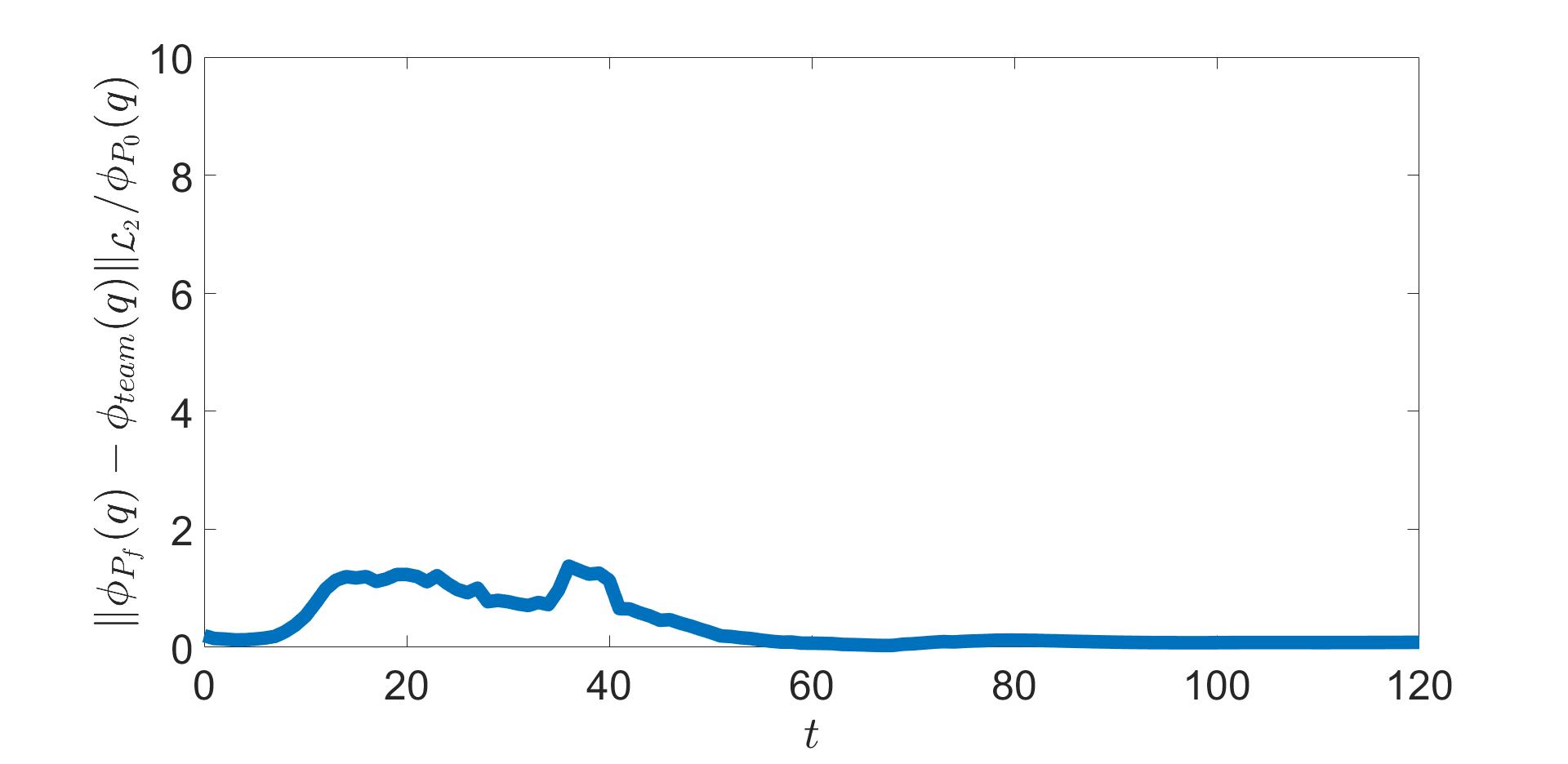}} 
	\caption{The time-evolution of the distance (measured by the spatial $\mathcal{L}_2$ norm) between the fine approximation of the team's density $\phi_{\mathrm{team}}(q,t)$ and the reference density $\phi(q,t)$ (Figure~\ref{fig5}(a)) and between $\phi_{\mathrm{team}}(q,t)$ and the final desired density $\phi_{P_f}(q)$ (Figure~\ref{fig5}(b)).} 
	\label{fig5}
\end{figure}

Fig.~\ref{fig5} shows the time-evolution of the distance in terms of the spatial $\mathcal{L}_2$ norm between the desired reference density distribution $\phi(q,t)$ and the fine approximation of the team's density distribution $\phi_{\mathrm{team}}(q,t)$ whose mean vectors, covariance matrices and mixing proportions are computed by using the GMMLO algorithm, which was implemented using the Fit Gaussian Mixture Model to Data function in MATLAB ($\mathrm{fitgmdist}(P(t),m)$). Also, the distance between the final desired density $\phi_{P_f}(q)$ and $\phi_{\mathrm{team}}(q,t)$ is shown in Fig.~\ref{fig5}. In both of these figures, we observe that the distance between the two densities decreases with time.
 
\section{Conclusion}
In this paper, a two-level method to the dynamic coverage distributed control problem of agents in an environment with obstacles was proposed. The aim of our approach is to steer a team of agents from an initial GM distribution to a terminal GM distribution while avoiding obstacles in the domain. First, we present a high-level approach corresponding to the closed-form GM density solution that provides a reference density path which is updated at each instant of time according to the agents' distribution. Second, we propose the solution to the low-level approach which corresponds to a distributed control law for the dynamic coverage problem that utilizes the GM solution of the high-level problem as a reference density in the optimization problem and guarantees collision avoidance in the cluttered domain while decreasing a locational cost defined over a (2-D non-autonomous cluttered domain). To guarantee obstacle avoidance in the low-level, we utilize a modified Voronoi tessellation that dynamically weights the boundaries between the obstacles and agents to create the OAVCs such that the latter are always tangent to the obstacles and never colliding. The control law presented is incorporated within a variation of Lloyd algorithm which utilizes the OAVCs. Thus, the agents are driven towards the time-varying centroids of their OAVCs to avoid collision with obstacles while tracking the desired reference coverage density. The approach proposed is verified through non-trivial simulations. This approach is limited to situations where the reference and the desired density distributions of the agents are known a priori and where the obstacles are static. Future research can be directed towards developing approaches for the deployment problem of multi-agent networks in dynamic environments and approaches to accommodate the avoidance of moving obstacles. Also, other approaches can be developed for estimating the reference density distribution overtime.

\ifCLASSOPTIONcaptionsoff
  \newpage
\fi

\bibliographystyle{IEEEtran}
\bibliography{ifacconf}

\begin{thebibliography}{10}
\providecommand{\url}[1]{#1}
\csname url@samestyle\endcsname
\providecommand{\newblock}{\relax}
\providecommand{\bibinfo}[2]{#2}
\providecommand{\BIBentrySTDinterwordspacing}{\spaceskip=0pt\relax}
\providecommand{\BIBentryALTinterwordstretchfactor}{4}
\providecommand{\BIBentryALTinterwordspacing}{\spaceskip=\fontdimen2\font plus
\BIBentryALTinterwordstretchfactor\fontdimen3\font minus
  \fontdimen4\font\relax}
\providecommand{\BIBforeignlanguage}[2]{{%
\expandafter\ifx\csname l@#1\endcsname\relax
\typeout{** WARNING: IEEEtran.bst: No hyphenation pattern has been}%
\typeout{** loaded for the language `#1'. Using the pattern for}%
\typeout{** the default language instead.}%
\else
\language=\csname l@#1\endcsname
\fi
#2}}
\providecommand{\BIBdecl}{\relax}
\BIBdecl

\bibitem{29}
A.~Pierson and D.~Rus, ``Distributed target tracking in cluttered environments
  with guaranteed collision avoidance,'' in \emph{2017 International Symposium
  on Multi-Robot and Multi-Agent Systems (MRS)}.\hskip 1em plus 0.5em minus
  0.4em\relax IEEE, 2017, pp. 83--89.

\bibitem{6}
J.~Cortes, S.~Martinez, T.~Karatas, and F.~Bullo, ``Coverage control for mobile
  sensing networks,'' \emph{IEEE Transactions on Robotics and Automation},
  vol.~20, no.~2, pp. 243--255, 2004.

\bibitem{10}
S.~Miah, M.~M.~H. Fallah, and D.~Spinello, ``Non-autonomous coverage control
  with diffusive evolving density,'' \emph{IEEE Transactions on Automatic
  Control}, vol.~62, no.~10, pp. 5262--5268, 2017.

\bibitem{13}
S.~Miah, A.~Y. Panah, M.~M.~H. Fallah, and D.~Spinello, ``Generalized
  non-autonomous metric optimization for area coverage problems with mobile
  autonomous agents,'' \emph{Automatica}, vol.~80, pp. 295--299, 2017.

\bibitem{27}
Y.-F. Chung and S.~Kia, ``A distributed service-matching coverage via
  heterogeneous mobile agents,'' September 2020.

\bibitem{37}
C.~Tabasso, V.~Cichella, S.~B. Mehdi, T.~Marinho, and N.~Hovakimyan,
  ``\BIBforeignlanguage{eng}{Time coordination and collision avoidance using
  leader-follower strategies in multi-vehicle missions},''
  \emph{\BIBforeignlanguage{eng}{Robotics (Basel)}}, vol.~10, no.~1, pp. 34--,
  2021.

\bibitem{47}
J.~Sun, J.~Tang, and S.~Lao, ``\BIBforeignlanguage{eng}{Collision avoidance for
  cooperative uavs with optimized artificial potential field algorithm},''
  \emph{\BIBforeignlanguage{eng}{IEEE Access}}, vol.~5, pp. 18\,382--18\,390,
  2017.

\bibitem{36}
Y.~Hu, H.~Yu, Y.~Zhong, and Y.~Lv, ``\BIBforeignlanguage{eng}{Distributed
  collision-avoidance formation control: A velocity obstacle-based approach},''
  in \emph{\BIBforeignlanguage{eng}{2019 IEEE Symposium Series on Computational
  Intelligence (SSCI)}}.\hskip 1em plus 0.5em minus 0.4em\relax IEEE, 2019, pp.
  1994--2000.

\bibitem{34}
P.~Ogren and N.~Leonard, ``\BIBforeignlanguage{eng}{A convergent dynamic window
  approach to obstacle avoidance},'' \emph{\BIBforeignlanguage{eng}{IEEE
  Transactions on Robotics}}, vol.~21, no.~2, pp. 188--195, 2005.

\bibitem{35}
D.~Zhou, Z.~Wang, S.~Bandyopadhyay, and M.~Schwager,
  ``\BIBforeignlanguage{eng}{Fast, on-line collision avoidance for dynamic
  vehicles using buffered voronoi cells},'' \emph{\BIBforeignlanguage{eng}{IEEE
  Robotics and Automation Letters}}, vol.~2, no.~2, pp. 1047--1054, 2017.

\bibitem{41}
A.~Pierson, W.~Schwarting, S.~Karaman, and D.~Rus, ``Weighted buffered voronoi
  cells for distributed semi-cooperative behavior,'' in \emph{2020 {IEEE}
  International Conference on Robotics and Automation, {ICRA} 2020, Paris,
  France, May 31 - August 31, 2020}.\hskip 1em plus 0.5em minus 0.4em\relax
  {IEEE}, 2020, pp. 5611--5617.

\bibitem{21}
A.~Okabe, B.~Boots, K.~Sugihara, and S.~N. Chiu, \emph{Spatial Tessellations :
  Concepts and Applications of Voronoi Diagrams}.\hskip 1em plus 0.5em minus
  0.4em\relax Hoboken: John Wiley and Sons, Incorporated, 2000.

\bibitem{19}
Q.~Du, M.~Emelianenko, and L.~Ju, ``Convergence of the {L}loyd algorithm for
  computing centroidal voronoi tessellations,'' \emph{SIAM Journal on Numerical
  Analysis}, vol.~44, no.~1, pp. 102--119, 2006.

\bibitem{22}
J.~Kennedy, A.~Chapman, and P.~M. Dower, ``Generalized coverage control for
  time-varying density functions,'' in \emph{2019 18th European Control
  Conference (ECC)}, June 2019, pp. 71--76.

\end{thebibliography}

\end{document}